\documentclass[referee]{aa} 
\input{psfig}

\def\parn{\par\noindent}
\def\Lp{L_{\rm p}}
\def\Reff{R_{\rm e}}
\def\xn{x_n}
\def\xnp{x_{n+1}}
\def\cn{c_n}
\def\Or{{\rm O}}
\def\Iz{I_0}
\def\Ieff{I_{\rm e}}
\def\fM{f_M}
\def\xiM{\xi_M}
\def\xir{\xi_r}
\catcode`\@=11
\def\gsim{\ifmmode{\mathrel{\mathpalette\@versim>}}
    \else{$\mathrel{\mathpalette\@versim>}$}\fi}
\def\lsim{\ifmmode{\mathrel{\mathpalette\@versim<}}
    \else{$\mathrel{\mathpalette\@versim<}$}\fi}
\def\@versim#1#2{\lower 2.9truept \vbox{\baselineskip 0pt \lineskip 
    0.5truept \ialign{$\m@th#1\hfil##\hfil$\crcr#2\crcr\sim\crcr}}}
\catcode`\@=12
\begin{document}

   \thesaurus{ (11.05.1; 
                11.06.2;  
                11.11.1;  
                11.16.1) }
   \title{Analytical properties of the R$^{1/m}$ law}

   \subtitle{}

   \author{L. Ciotti\inst{1,2}
           \and
           G. Bertin\inst{2}}

   \offprints{L. Ciotti}

   \institute{Osservatorio Astronomico di Bologna, via Ranzani 1,
              I-40127 Bologna\\
              email: ciotti@bo.astro.it
              \and              
              Scuola Normale Superiore, Piazza dei Cavalieri 7,
              I-56126 Pisa\\
              email: bertin@sns.it}

   \date{Accepted}

   \maketitle

   \begin{abstract}

In this paper we describe some analytical properties of the $R^{1/m}$
law proposed by Sersic (1968) to categorize the photometric profiles
of elliptical galaxies.  In particular, we present the full asymptotic
expansion for the dimensionless scale factor $b(m)$ that is introduced
when referring the profile to the standard effective radius.
Surprisingly, our asymptotic analysis turns out to be useful even for
values of $m$ as low as unity, thus providing a unified analytical
tool for observational and theoretical investigations based on the
$R^{1/m}$ law for the entire range of interesting photometric
profiles, from spiral to elliptical galaxies.

      \keywords{Galaxies: elliptical and lenticular, cD --
                Galaxies: fundamental parameters --
                Galaxies: kinematics and dynamics --
                Galaxies: photometry
               }
   \end{abstract}

%

\section{Introduction}

After its introduction as a generalization of the $R^{1/4}$ law (de
Vaucouleurs 1948), the so--called Sersic law (Sersic 1968) has found a
variety of applications. On the observational side, it has been used
as a tool to quantify the non-homology of elliptical galaxies (see,
e.g., Davies et al. 1988; Capaccioli 1989, hereafter C89; Caon,
Capaccioli \& D'Onofrio 1993; Young \& Currie 1994; D'Onofrio,
Capaccioli \& Caon 1994; Prugniel \& Simien 1997, hereafter PS97;
Wadadekar, Robbason \& Kembhavi 1999). In addition, it has been
applied to the description of the the surface brightness profiles of
galaxy bulges (see Andredakis, Peletier \& Balcells 1995; Courteau, De
Jong \& Broeils 1996). One research area where the usefulness of the
Sersic law as a statistically convenient description has been
exploited is that of the Fundamental Plane of elliptical galaxies
(Graham et al. 1996; Ciotti, Lanzoni \& Renzini 1996; Graham \&
Colless 1997; Ciotti \& Lanzoni 1997; Graham 1998). On the theoretical
side, it has been the focus of several general investigations (see,
e.g., Makino, Akiyama \& Sugimoto 1990; Ciotti 1991, hereafter C91;
Gerbal et al. 1997; Andredakis 1998).

According to such law, the surface brightness profile is given by
\begin{equation}
I(R)=\Iz e^{-b\eta^{1/m}},
\end{equation}
where $\eta=R/\Reff$, $m$ is a positive real number, and $b$ a
dimensionless constant such that $\Reff$ is the effective radius,
i.e., the projected radius encircling half of the total luminosity
associated with $I(R)$.  In a broad statistical sense, it is found
that bright ellipticals are well fitted by the Sersic law with $m$
around 4, dwarf ellipticals and galaxy disks with $m$ around 1, and
finally bulges and intermediate luminosity ellipticals with $1\leq
m\leq 4$. For some galaxies, a value of $m$ even higher than 10 has
been found (e.g., see NGC 4552, Caon et al. 1993).

The projected luminosity inside the projected
radius $R$ is given by
\begin{equation}
L (R)=2\pi\int_0^RI(R')R'dR'=
        \Iz\Reff^2 {2\pi m\over b^{2m}}\gamma (2m,b\eta ^{1/m}),
\end{equation}
where (for $\alpha >0$)
\begin{equation}
\gamma(\alpha,x)=\int_0^xe^{-t}t^{\alpha-1}dt
\end{equation}
is the \emph{(left) incomplete gamma function\/}. The total luminosity is
then given by
\begin{equation}
L=\Iz\Reff^2 {2\pi m\over b^{2m}}\Gamma (2m),
\end{equation}
where $\Gamma(\alpha)=\gamma(\alpha,\infty)$ is the \emph{complete
gamma function\/}.  From the definition of $\Reff$ it follows that
$b(m)$ is the solution of the following equation:
\begin{equation}
\gamma (2m,b) ={\Gamma (2m)\over 2}.
\end{equation}

\section{Asymptotic expansion}
Unfortunately, Eq. (5) cannot be solved in explicit, closed form, and
so it is usually solved numerically\footnote{For $m=1$, i.e., the
exponential profile, the solution can be formally expressed using the
\emph{Lambert\/} $W$ function, as
$b(1)=-1-W(-1,-1/2e)=1.678346990...$}.  This is inconvenient for a
number of observational and theoretical applications.  The exact
values of $b(m)$ are recorded in Table 1 for $1\leq m\leq 10$.  For
the de Vaucouleurs law, $m=4$ and $b(4)\approx 7.66924944$.
Interpolation formulae for $b(m)$ have been given in the literature,
namely $b\simeq 1.9992m-0.3271$ by C89 (as reported by Graham \&
Colless 1997), $b\simeq 2m-0.324$ by C91, $b\simeq 2m - 1/3$ (for $m$
integer) by Moriondo, Giovanardi \& Hunt (1998), and the ``numerical
solution'' $b(m)\simeq 2m -1/3 +0.009876/m$ by PS97.  These
expressions provide an accurate fit in the range $0.5\leq m\leq 10$;
curiously, their leading term is \emph{linear\/} in $m$, with a slope
very close to 2.  In the following we show that this behavior results
from a general property of the gamma function.

Prompted by Eq. (5) we now address the following:\parn
{\bf Problem} {\it Solve for $x$
\begin{equation}
\gamma(\alpha,x)={\Gamma(\alpha)\over 2},
\end{equation}
for given $\alpha>0$.}

Because no explicit solution in closed form is available, we will
focus on the asymptotic expansion of $x(\alpha)$ for $\alpha >>1$. In
fact, it is well known that in many cases asymptotic expansions turn
out to give excellent approximations of the true function even for
relatively small values of the expansion parameter.  The starting
point of our study is the asymptotic relation (see Abramowitz \&
Stegun 1965)
\begin{equation}
\Gamma(\alpha)\sim e^{-\alpha}\alpha^{\alpha}\sqrt{{2\pi\over\alpha}}
\left[1+{1\over 12\alpha}+
                           {1\over 288\alpha^2}-
                           {139\over 51840\alpha^3}-
                           {571\over 2488320\alpha^4}+
                           {163879\over 209018880\alpha^5}+
                           \Or(\alpha^{-6})\right].
\end{equation}
This is the \emph{Stirling formula\/}\footnote{The derivation of this
formula can be found in standard textbooks.  The coefficients
appearing in the asymptotic expansion of $\ln\Gamma(\alpha)$ for
$\alpha\to\infty$ can be expressed in terms of the so--called
\emph{Bernoulli numbers\/}; see, e.g., Arfken \& Weber 1995, Chapts. 5
and 10.}, which is known to be associated with a relative error
smaller than $3\times 10^{-6}$ already for $\alpha=2$.

Let us now introduce the sequence
\begin{equation}
\xn =\alpha+\sum_{k=0}^{n-1}{c_k\over\alpha^k}
\end{equation}
with $x_0=\alpha$, so that $\xnp=\xn+c_n/\alpha^n$. Here $c_k$ are
coefficients (to be determined at a later stage), independent of
$\alpha$. Then we start by proving the following asymptotic results,
applicable for $\alpha >>1$.
\parn
{\bf Lemma 1} {\it The following asymptotic relation
\begin{equation}
\gamma(\alpha,x_0)\sim{\Gamma(\alpha)\over 2}+
                           e^{-\alpha}\alpha^{\alpha}\sum_{k=0}^{\infty}
                           {P_k^{(0)}\over\alpha^{k+1}},
\end{equation}
holds, where $P_k^{(0)}$ are rational numbers.}
\parn
The validity of Eq. (9) can be established by means of a standard
asymptotic expansion (e.g., see Bender \& Orszag 1978, Bleinstein \&
Handelsman 1986) of the integral
\begin{equation}
\gamma(\alpha,\alpha)=\alpha^{\alpha}e^{-\alpha}\int_{-1}^0
                        {\exp[-\alpha s+\alpha\ln(1+s)]\over 1+s}ds.
\end{equation} 
In fact, the argument of the integral is the same as that of the
integral representation of $\Gamma(\alpha)$. In both cases the
stationary point for the exponent occurs at $s=0$, but for
$\Gamma(\alpha)$ the stationary point is in the middle of the domain
of integration, because the integral extends to $\infty$ (instead, for
the integral in Eq. [10] the upper limit is precisely $s=0$). Thus,
when we consider the power series expansion (in $s$) of the argument
of the integral around the stationary point, for
$\gamma(\alpha,\alpha)$ the \emph{even\/} powers of $s$ contribute
exactly \emph{one half\/} of their contribution to $\Gamma(\alpha)$,
while the \emph{odd\/} powers determine the terms in Eq. (9) associated
with the coefficients $P^{(0)}_k$ (in contrast, the odd powers do not
contribute to $\Gamma(\alpha)$, by symmetry). The calculation of
$P^{(0)}_k$ is tedious, but straightforward.

Note that there is a ``shift'' of powers, by $\alpha^{1/2}$, between
the two terms on the right hand side of Eq. (9). In particular, the
second term is smaller by a factor $\Or(\alpha^{-1/2})$. This already
shows that $x_0=\alpha$ is a first approximate solution to the problem
set by Eq. (6).
\parn
{\bf Lemma 2} {\it The following asymptotic relation
\begin{equation}
\gamma(\alpha,\xnp)\sim\gamma(\alpha,\xn)+e^{-\alpha}\alpha^{\alpha}
                         f(\alpha)
\end{equation}
holds, with $f(\alpha)=\Or(\alpha^{-n-1})$. To leading order, $f(\alpha)\sim
c_n/\alpha^{n+1}$.}
\parn
This result easily follows from the definitions of the 
quantities involved (Eqs. [3] and [8]), which give
\begin{equation} 
\gamma(\alpha,\xnp)=\gamma(\alpha,\xn)+
                      e^{-\xn}\int_0^{\cn/\alpha^n}
                      e^{-t}(t+\xn)^{\alpha -1}dt.
\end{equation}

At this point we can proceed to prove the following theorem: \parn
{\bf Theorem} {\it For large (real) values of $\alpha$, the full
asymptotic expansion of the solution to the problem posed by Eq. (6)
can be expressed as
\begin{equation}
x(\alpha)=\alpha+\sum_{n=0}^{\infty}{\cn\over\alpha^n},
\end{equation}
where
\begin{equation}
\cn=-P_n^{(n)},
\end{equation}
and the coefficients $P_k^{(n)}$ can be calculated by {\rm iteration}
on the relation}
\begin{equation}
\gamma(\alpha,\xn)\sim\Gamma(\alpha)/2+e^{-\alpha}\alpha^{\alpha}
                                         \sum_{k=n}^{\infty}
                              {P_k^{(n)}\over\alpha^{k+1}}.
\end{equation}
The proof is obtained by induction. In fact, Eq. (9) shows that the
statement is true for $n=0$, with the coefficients $P^{(0)}_k$
available from the asymptotic analysis outlined in the proof of Lemma
1. If we now refer to the result of Lemma 2, with the leading order
expression for $f(\alpha)$, and assume the statement (related to
Eq.[15]) to hold true for $\xn$, we find
\begin{equation} 
\gamma(\alpha,\xnp)\sim{\Gamma(\alpha)\over 2}+e^{-\alpha}\alpha^{\alpha}
                         \left[\sum_{k=n}^{\infty}
                               {P_k^{(n)}\over\alpha^{k+1}}+
                               {\cn\over\alpha^{n+1}} + ...
                         \right].
\end{equation}
In other words, the statement is found to hold true also for $\xnp$,
provided $\cn=-P_n^{(n)}$, as required by Eq. (14). The method thus
provides a way to systematically improve the approximation to
$x(\alpha)$ by means of $n$ steps, leading to an estimate $\xn$; step
by step the possible presence of undesidered ``shifted'' (odd) terms
is eliminated and the process leads to the complete determination of
the coefficients defining the asymptotic series (13). At a given level
$n$ of desidered accuracy, the coefficients $P^{(n)}_k$ depend on the
values $P^{(i)}_k$ for $i=0,...,n-1$.

The explicit computation yields: 
\begin{equation}
x(\alpha)\sim\alpha-{1\over 3}
              +{8\over 405\alpha}
              +{184\over 25515\alpha^2}
              +{1048\over 1148175\alpha^3}
              -{17557576\over 15345358875\alpha^4}
              +\Or(\alpha^{-5}).
\end{equation}
The first two terms can be easily checked using standard general
formulae for the leading terms of the relevant steepest descent
asymptotic expansion.

\section{Analytical properties of the Sersic law}

Therefore, the first terms of the asymptotic expansion of $b(m)$ 
(for real $m$) are
\begin{equation}
b(m)\sim    2m-{1\over 3}
              +{4\over 405 m}
              +{46\over 25515 m^2}
              +{131\over 1148175 m^3}
              -{2194697\over 30690717750 m^4}
              +\Or(m^{-5}).
\end{equation}
Equation (18) now clearly explains the value of the interpolation
formulae found earlier (C89, C91, PS97); note that
$4/405=0.0098765...$.  How many terms in the asymptotic expansion are
required to obtain a better representation of $b(m)$ when compared to
the previously introduced interpolations?

We have computed the relative errors of the various expressions with
respect to the true value of $b(m)$ (obtained by solving numerically
Eq. [5] with a precision of 20 significant digits) for integer values
of $m$ in the range $1\leq m\leq 10$, and the results are reported in
Table 1.  The first result is that using the first four terms of the
expansion the true value of $b(m)$ is recovered with a relative error
of $6\times 10^{-7}$ for $m=1$, and $4\times 10^{-9}$ for $m=10$,
i.e., the asymptotic expansion so truncated performs much better than
the formulae cited previously.  Obviously, for larger values of $m$
the error becomes correspondingly smaller.  The second somewhat
surprising result is the fact that Eq. (18) is already very accurate
for $m$ as small as 1.  This allows us to include, within the reach of
the present analysis, the case of exponential profiles.  A third point
that we have noted is that, for \emph{fixed\/} $m$, there is an \emph{
optimal truncation\/} of the asymptotic expansion, beyond which, as is
well known in the general context of asymptotic analysis, increasing
the number of terms in the expansion does not improve the accuracy of
the estimate.  For example, for $m=1$, the optimal truncation occurs
at the fourth term, for which the attained relative error is $6\times
10^{-7}$.
   \begin{table*} \caption[]{True values and relative errors on $b(m)$
      for integer values of $m$, using the C89, C91, and PS97
      formulae. As(4) is the relative error for the asymptotic
      expansion given in Eq. (18), truncated to the first four terms.}
      \label{Tab1} 
\halign{\strut# & #\hfil &\quad\hfil#\hfil
      &\quad\hfil#\hfil &\quad\hfil#\hfil &\quad\hfil#\hfil
      &\quad\hfil#\hfil &\quad\hfil#\cr \noalign{\hrule \vskip 5 pt} &
      $m$ & $b(m)$ & C89 & C91 & PS97 & As(4) \cr \noalign{\vskip 5 pt
      \hrule \vskip 5 pt} & 1 & 1.67834699 &$4\times
      10^{-3}$&$10^{-3}$&$10^{-3}$&$6\times 10^{-7}$&\cr & 2 &
      3.67206075 &$2\times 10^{-4}$&$10^{-3}$&$10^{-4}$&$10^{-6}$&\cr
      & 3 & 5.67016119 &$6\times 10^{-5}$&$10^{-3}$&$4\times
      10^{-5}$&$4\times 10^{-7}$&\cr & 4 & 7.66924944 &$6\times
      10^{-5}$&$9\times 10^{-4}$&$10^{-5}$&$10^{-7}$&\cr & 5 &
      9.66871461 &$2\times 10^{-5}$&$8\times 10^{-4}$&$8\times
      10^{-6}$&$5\times 10^{-8}$&\cr & 6 & 11.6683632 &$2\times
      10^{-5}$&$7\times 10^{-4}$&$4\times 10^{-6}$&$3\times
      10^{-8}$&\cr & 7 & 13.6681146 &$6\times 10^{-5}$&$6\times
      10^{-4}$&$3\times 10^{-6}$&$2\times 10^{-8}$&\cr & 8 &
      15.6679295 &$9\times 10^{-5}$&$5\times 10^{-4}$&$2\times
      10^{-6}$&$9\times 10^{-9}$&\cr & 9 & 17.6677864
      &$10^{-4}$&$5\times 10^{-4}$&$10^{-6}$&$6\times 10^{-9}$&\cr &
      10& 19.6676724 &$10^{-4}$&$4\times 10^{-4}$&$9\times
      10^{-7}$&$4\times 10^{-9}$&\cr \noalign{\vskip 5 pt \hrule
      \vskip 2 pt}\cr}
\end{table*}
For simplicity, in the following part of this Section we will record a
number of interesting analytical expressions restricted to their
leading order. Of course, the asymptotic analysis provided here would
allow us to give explicitly any higher order term, not shown below, if
so desired.

\subsection{Total luminosity and central potential for a spherical system 
             with an R$^{1/m}$ projected luminosity profile}
From Eqs. (4) and (18) the total luminosity is found to be 
\begin{equation}
L=\Iz\Reff^2 {2\pi m\over b^{2m}}\Gamma (2m)\sim
    \Iz\Reff^2 2\pi^{3/2}e^{1/3}e^{-2m}\sqrt{m}[1+\Or(m^{-1})],
\end{equation}
where we have used the fact that 
$b^{2m}\sim e^{-1/3}(2m)^{2m}[1+\Or(m^{-1})]$. 
Following C91, the central potential of the spherically symmetric
density distribution associated with the Sersic law is given by:
\begin{equation}
\Phi_0=-G{M\over L}\Iz\Reff{4\Gamma (1+m)\over b^m}.
\end{equation}
Here $M$ is the total mass of the system, and the mass--to--light
ratio is taken to be constant. Thus, an asymptotic estimate at given
$\Iz$, $\Reff$ is
\begin{equation}
\Phi_0\sim -{GM\over L}\Iz\Reff 2^{5/2}\pi^{1/2}e^{1/6}
              {\sqrt{m}\over (2e)^m}[1+\Or(m^{-1})].
\end{equation}
We note that, for $m=1$ and $m=4$, a truncation of Eq. (19) to the
leading term is characterized by a relative error of 5.7 per cent and
of 1.5 per cent, respectively. The corresponding truncation on
Eq. (21) is associated with a relative error of 8.6 per cent and 2.3
per cent.

The proper normalization of the R$^{1/m}$ profile, to be
considered for a case with given scales $L$ and $\Reff$, is
\begin{equation} 
I(R)={L\over\Reff^2}{b^{2m}e^{-b\eta^{1/m}}\over 2\pi m\Gamma(2m)},
\end{equation}
which thus provides the useful quantity
\begin{equation} 
\Ieff=I(\eta=1)\sim {L\over\Reff^2}{1\over 2\pi^{3/2}\sqrt{m}}
        [1+\Or(m^{-1})],
\end{equation}
so that
\begin{equation}
\Phi_0\sim -{GM\over\Reff}{2^{3/2}\over\pi e^{1/6}}
              \left({e\over 2}\right)^m[1+\Or(m^{-1})].
\end{equation}
For $m=1$ and $m=4$, a truncation of Eq. (23) to the leading term is
characterized by a relative error of 7.3 per cent and of 1.8 per cent,
respectively.  The corresponding truncation on Eq. (24) is associated
with a relative error of 3.1 per cent and 0.8 per cent.

\section{Conclusions}
In this paper, the full asymptotic expansion for the dimensionless
scale factor $b(m)$ appearing in the Sersic profile has been
constructed.  It is shown that this expansion, even when truncated to
the first four terms as
\begin{equation}
b(m)=2m-{1\over 3}+{4\over 405 m}+{46\over 25515 m^2}
\end{equation}
performs much better than the formulae given by C89, C91 and PS97,
even for $m$ values as low as unity, with relative errors smaller than
$\simeq 10^{-6}$. The use of this simple formula is thus recommended
both in theoretical and observational investigations based on the
Sersic law. With the aid of this formula, we have been able to clarify
a number of interesting properties associated with the Sersic profile.
The additional material presented in Appendix A can be compared to the
simple power law $R^{-2}$, often used in the past to fit the
photometric profiles of elliptical galaxies.

\begin{acknowledgements}
We would like to thank F. Simien for several useful suggestions.  This
work was supported by MURST, contract CoFin98.  L.C. was also
partially supported by ASI, contract ASI-ARS-96-70.
\end{acknowledgements}

\appendix
\section{Remarks on a power law approximation}
The surface brightness profile given in Eq. (1) is sometimes expressed as
(see Eqs. [22]-[23])
\begin{equation}
I(R)=\Ieff e^{b(1-\eta^{1/m})}.
\end{equation}
In this case,
for a \emph{fixed\/} location $\eta>0$, we may consider an asymptotic
expansion of the surface brightness profile at constant $\Ieff$ for
$m>>1$. This is obtained by introducing a stretched radial coordinate
$\xi=\ln\eta$ and by noting that
\begin{equation}
b(1-\eta^{1/m})=b(1-e^{\xi/m})=
-2\xi\left(1-{1\over 6m}+{2\over 405m^2}\, \ldots\right)
\left(1+{\xi\over 2m}+{\xi^2\over 6m^2}\, \ldots\right).
\end{equation}
Thus, we find
\begin{equation}
I(R)\sim{\Ieff\over\eta^2}[1+\Or(m^{-1})].
\end{equation}
We may recall here that the photometric profiles of elliptical
galaxies have often been described in the past in terms of power laws
(see, e.g., Hubble 1930). A naive inspection of the first term omitted
in the expansion (A2) suggests that Eq. (A3) is adequate provided
$|\xi (\xi -1/3)/m| <<1$. Note that the term involved vanishes at
$\xi=0$ and at $\xi=1/3$. This is an indication that the quality of
the power law approximation is asymmetric, with a modest bias to the
outer region.
\begin{figure}
\parbox{1cm}{
\psfig{file=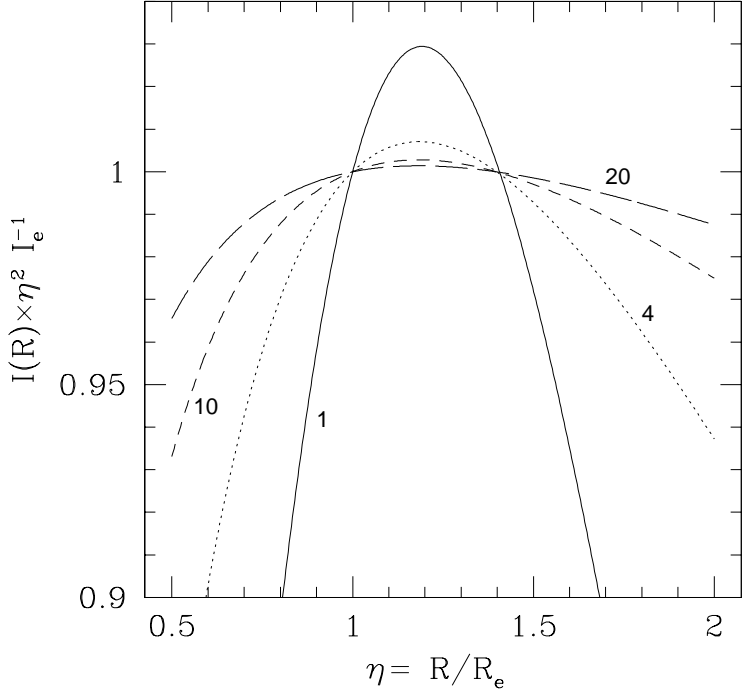,width=13cm,height=13cm,angle=0}}
\caption[]{Ratio between the $R^{1/m}$ profile (normalized at $\Ieff$) and
           a properly scaled $R^{-2}$ profile, for $m=1,4,10,20$.}
\label{Fig1}
\end{figure}

Consider the function $f(\xi)=b(1-e^{\xi/m})+2\xi$.  The quantity
$\exp{(f)}$ gives the ratio between the $R^{1/m}$ profile and its
$R^{-2}$ approximation. The function $f$ diverges to $-\infty$ both
for $\xi\to -\infty$ (i.e., $\eta\to 0$) and for $\xi\to +\infty$
(i.e., $\eta\to +\infty$), and has a single maximum $\fM$ at $\xiM$,
defined by the relation $e^{\xiM/m}=2m/b$, where
\begin{equation}
\fM=b-2m+2m\,\ln\left({2m\over b}\right)\sim {1\over 36m}+\Or(m^{-2}).
\end{equation}
Note that $2m/b$ is close to unity (in fact, $\xiM\sim 1/6$), i.e.,
that $\xiM/m$ is small.  Therefore, the power law profile is slightly
underluminous with respect to the $R^{1/m}$ profile in the radial
range between the effective radius and an outer radius $\xir\sim
2\xiM\sim 1/3$, where $f(\xir)=0$, which coincides with the outer
location identified by a previous naive inspection (see comment after
Eq. [A3]). Outside such radial range the power law profile is brighter
than the $R^{1/m}$ profile. In such region, where $f<0$, we may ask
how far (in radial range) Eq. (A3) applies, by studying the condition
$1-e^f\leq\epsilon$, i.e., $|f|\leq\epsilon$.  We expand $f$ for
negative values of $\xi$ around $\xi=0$, and around $\xir$ for $\xi
>\xir$. The range of applicability of Eq. (A3) is thus constrained by
the condition $(1-3\epsilon)^m\lsim\eta\lsim (1+3\epsilon)^me^{1/3}$.
This situation is illustrated in Fig. 1.

\end{document}